\documentclass[]{emulateapj}

\shorttitle{Stellar Activity Planet Surface Gravity Correlation}
\shortauthors{Hartman}

\begin{document}

\title{A Correlation Between Stellar Activity and the Surface Gravity of Hot Jupiters}
\author{J.~D.~Hartman\altaffilmark{1}}
\altaffiltext{1}{Harvard-Smithsonian Center for Astrophysics, 60 Garden St., Cambridge, MA~02138, USA; jhartman@cfa.harvard.edu;}

\begin{abstract}
Recently \citet{Knutson.10} have demonstrated a correlation between
the presence of temperature inversions in the atmospheres of hot
Jupiters, and the chromospheric activity levels of the host
stars. Here we show that there is also a correlation, with greater
than 99.5\% confidence, between the surface gravity of hot Jupiters
and the activity levels of the host stars, such that high surface
gravity planets tend be found around high activity stars. We also find
a less significant positive correlation between planet mass and
chromospheric activity, but no significant correlation is seen between
planet radius and chromospheric activity. We consider the possibility
that this may be due to an observational bias against detecting lower
mass planets around higher activity stars, but conclude that this bias
is only likely to affect the detection of planets much smaller than
those considered here. Finally, we speculate on physical origins for
the correlation, including the possibility that the effect of stellar
insolation on planetary radii has been significantly underestimated,
that strong UV flux evaporates planetary atmospheres, or that high
mass hot Jupiters induce activity in their host stars, but do not find
any of these hypotheses to be particularly compelling.
\end{abstract}


\section{Introduction}\label{sec:intro}

With more than 70 transiting exoplanets (TEPs) now
known\footnote{e.g. http://exoplanets.org}, it has been become possible
to detect statistically robust correlations between the parameters of
TEPs and their host stars, which in turn yields insights into the
processes that are important for determining the physical properties
of exoplanet systems. Several correlations have already been noted,
including correlations between the masses and orbital periods of TEPs
\citep{Gaudi.05, Mazeh.05,2008ApJ...677.1324T}, between their surface
gravities and orbital periods
\citep{Southworth.07,2008ApJ...677.1324T}, between the inferred core
mass of planets and the metallicity of their host stars
\citep{Guillot.06, Burrows.07}, between Safronov number and the host
star metallicity \citep{2008ApJ...677.1324T}, and between the radii of
planets and their average equilibrium temperature and host metallicity
\citep{Enoch.10}.

Very recently \citet[][hereafter KHI10]{Knutson.10} have demonstrated
a correlation between the emission spectra of TEPs and the
chromospheric activity levels of their host stars, as measured from
the strength of the emission lines at the Ca~II H and K line
cores. Planets with spectra consistent with noninverted temperature
models appear to be found around high activity stars, while planets
with spectra consistent with temperature inversions are found around
low activity stars. In demonstrating this correlation KHI10 also
published a catalogue of $\log R^{\prime}_{HK}$ values for 39
TEPs. This new, homogeneous sample enables statistical studies of the
relationships between stellar activity and the physical properties of
TEPs.

In this paper we use the sample of $\log R^{\prime}_{HK}$ values from
KHI10 to investigate correlations between stellar activity and other
TEP parameters. We find that there is a significant correlation
between $\log R^{\prime}_{HK}$ and the planet surface gravity $\log
g_{\rm P}$. A similar correlation between stellar activity (as traced
by the temporal variation in an index related to $\log
R^{\prime}_{HK}$) and the minimum planetary mass $M_{\rm P}\sin i$ was
previously noted by \citet{Shkolnik.05} for a sample of 10 RV planets,
though the authors deemed the correlation to be only suggestive. Here
we demonstrate that the $\log R^{\prime}_{HK}$-$\log g_{\rm P}$
correlation is robust with greater than $99.5\%$ confidence.

The structure of this paper is as follows: in
Section~\ref{sec:dataandanal} we describe the data and conduct the
statistical analysis to establish the correlation, in
Section~\ref{sec:selecteffect} we discuss a potential observational
bias which might lead to this correlation, and in
Section~\ref{sec:discussion} we speculate on the physical origins of
this correlation.

\section{Data and Statistical Analysis}\label{sec:dataandanal}

Table~\ref{tab:rhkvslogg} gives the $\log(R^{\prime}_{\rm HK})$, and
$\log g_{P}$ values adopted for planets with both parameters measured,
together with the sources from the literature for the surface gravity
measurements. In all cases we take $\log(R^{\prime}_{\rm HK})$ from
KHI10. In general we take the surface gravity of planets from studies
which calculated it directly from observable parameters \citep[the
  transit duration, depth, and impact parameter, together with the RV
  semiamplitude, eccentricity, and orbital period;
  see][]{Southworth.07} in a Markov-Chain Monte Carlo simulation, or
we calculate it ourselves from the given observable parameters.

\begin{figure}[ht]
\epsscale{1.2} \plotone{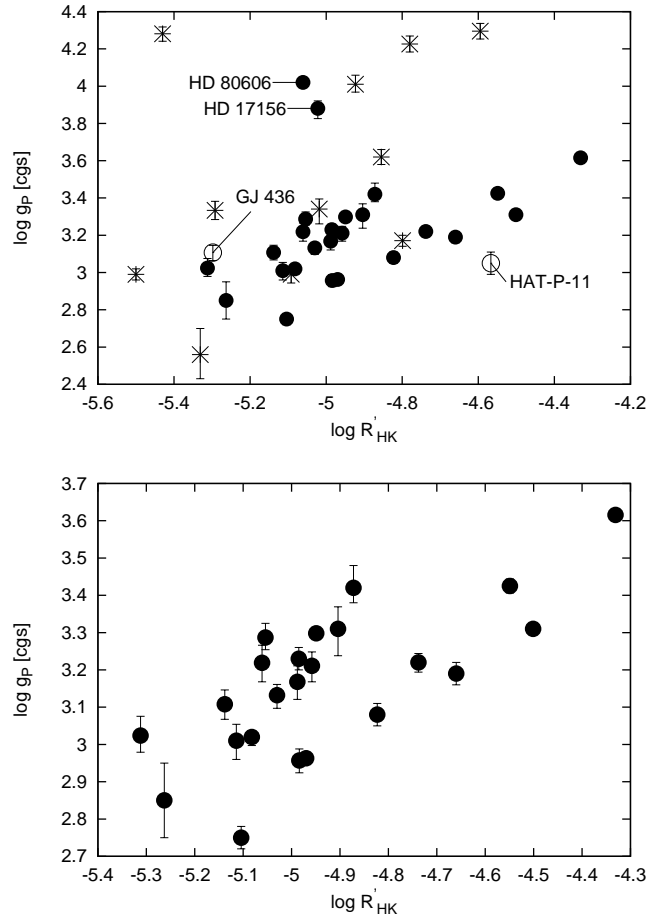}
\caption{Top: Surface gravity of transiting exoplanets vs. the
  chromospheric activity of the stellar hosts, as measured with $\log
  R^{\prime}_{\rm HK}$. The values and data sources are given in
  Table~\ref{tab:rhkvslogg}. Filled circles show planets with $M >
  0.1~M_{\rm J}$ orbiting host stars with $4200~{\rm K} < T_{\rm eff}
  < 6200~{\rm K}$, crosses show planets with $M > 0.1~M_{\rm J}$
  orbiting host stars with $T_{\rm eff} < 4200~{\rm K}$ or $T_{\rm
    eff} > 6200~{\rm K}$ (i.e. outside the range over which $\log
  R^{\prime}_{\rm HK}$ is calibrated), and open circles show two hot
  Neptunes with $M < 0.1~M_{\rm J}$, which might be expected to have
  different atmospheric properties than more massive planets. We also
  label the two planets HD~80606b and HD~17156b, with semi-major axes
  $a > 0.1~{\rm AU}$, whose properties are less likely to be
  influenced by the stellar flux. These planets have relatively high
  surface gravities most likely as a result of reduced stellar
  insolation. Bottom: Same as above, here we only show planets with $M
  > 0.1~M_{\rm J}$, $a < 0.1~{\rm AU}$ orbiting stars with $4200~{\rm
    K} < T_{\rm eff} < 6200~{\rm K}$.}
\label{fig:rhkvslogg}
\end{figure}

Figure~\ref{fig:rhkvslogg} shows the relation between $\log
R^{\prime}_{\rm HK}$ and $\log g_{\rm P}$; the existence of a
correlation is readily apparent. To determine the statistical
significance of this correlation we use the Spearman rank-order
correlation test \citep[see][]{Press.92}, finding $r_{S} = 0.45$. The
probability that a random sample of size $N = 39$ drawn from an
uncorrelated population would have either $r_{S} >= 0.45$ or $r_{S} <=
-0.45$ is only $0.4\%$, so the significance of the correlation is
99.6\%. If we exclude the two hot Neptunes GJ~436b and HAT-P-11b,
which one might expect to have atmospheric properties that are
different from more massive planets, the two long period planets
HD~80606b and HD~17156b, and also exclude planets orbiting stars which
have temperatures outside the range over which $\log R^{\prime}_{\rm
  HK}$ has been calibrated \citep[$4200~{\rm K} < T_{\rm eff} <
  6200~{\rm K}$;][]{Noyes.84}, we find a stronger correlation of
$r_{S} = 0.68$, with a false alarm probability of 0.032\%. The sample
size in this case is $N = 23$.

We have also searched for correlations between $\log R^{\prime}_{\rm
  HK}$ and other parameters such as planetary mass, planetary radius,
the radial velocity semi-amplitude, stellar mass, planetary density,
the Safronov number $\theta$ \citep{Hansen.07}, planetary equilibrium
temperature (assuming zero albedo), average stellar flux
incident on the planet, orbital period, stellar metallicity, and
stellar effective temperature. Table~\ref{tab:correlationcoeffs}
summarizes the strength of each correlation for three separate
samples:
\begin{enumerate}
\item $M > 0.1~M_{\rm J}$, $a < 0.1~{\rm AU}$, and $4200~{\rm K} < T_{\rm eff} < 6200~{\rm K}$.
\item $4200~{\rm K} < T_{\rm eff} < 6200~{\rm K}$
\item No restrictions
\end{enumerate}

In addition to the $\log R^{\prime}_{\rm HK}$-$\log g_{\rm P}$
correlation, positive correlations with $> 99\%$ confidence are also
seen between $\log R^{\prime}_{\rm HK}$ and $\theta$, and between
$\log R^{\prime}_{\rm HK}$ and $\rho_{\rm P}$. All three parameters
($\log g_{\rm P}$, $\theta$ and $\rho_{\rm P}$) scale as $M_{\rm
  P}/R_{\rm P}^{n}$ ($n = 1$ for $\theta$, $n = 2$ for $\log g_{\rm
  P}$ and $n = 3$ for $\rho_{\rm P}$). The correlations seen between
these parameters and $\log R^{\prime}_{\rm HK}$ most likely have the
same origin. We focus on $\log g_{\rm P}$ here both because the
correlation is slightly more significant for this parameter than it is
for $\theta$ or $\rho_{\rm P}$, and because for TEPs $\log g_{\rm P}$
can be determined directly from measurable parameters
\citep{Southworth.07}, while the other two parameters are dependent on
stellar models, which could conceivably introduce a bias if there is a
systematic error in the stellar models which depends on activity.

By checking several different parameters for correlations with $\log
R^{\prime}_{\rm HK}$, we have conducted several independent trials, and
must therefore increase the false alarm probabilities to account for
this. Several of the parameters are strongly correlated ($\log g_{\rm
P}$, $\theta$ and $\rho_{\rm P}$, as well as $K$ and $M_{\rm P}$, and
$T_{\rm eq,P}$ and $\langle F \rangle_{\rm P}$), so these are not
completely independent trials; we estimate that the total number of
independent trials is between $n = 6$ and $n = 12$. The corrected
false alarm probability for the $\log g_{\rm P}$-$\log R^{\prime}_{\rm
HK}$ correlation is $1 - (1 - 0.00032)^n = 0.19\%$ or $0.38\%$ for $n
= 6$ and $12$ respectively.

We note from table~\ref{tab:correlationcoeffs} that while the planet
mass $M_{\rm P}$ shows a positive correlation with $\log
R^{\prime}_{\rm HK}$ with $\sim 97\%$ confidence, there is no
significant correlation detected between $\log R^{\prime}_{\rm HK}$
and the planet radius $R_{\rm P}$.

\begin{deluxetable}{lrrrrrr}
\tabletypesize{\footnotesize}
\tablewidth{0pc}
\tablecaption{Correlation Between $\log{R^{\prime}_{\rm HK}}$ and Other Parameters}
\tablehead{
&
\multicolumn{2}{c}{Sample 1} &
\multicolumn{2}{c}{Sample 2} &
\multicolumn{2}{c}{Sample 3} \\
\colhead{Parameter} &
\colhead{$r_{S}$\tablenotemark{a}} &
\colhead{FAP\tablenotemark{b}} &
\colhead{$r_{S}$\tablenotemark{a}} &
\colhead{FAP\tablenotemark{b}} &
\colhead{$r_{S}$\tablenotemark{a}} &
\colhead{FAP\tablenotemark{b}} 
}
\startdata
$\log g_{\rm P}$ & 0.68 & 0.032\% & 0.47 & 1.5\% & 0.45 & 0.39\% \\
$\theta$ & 0.62 & 0.17\% & 0.36 & 7.0\% & 0.41 & 0.91\% \\
$\rho_{\rm P}$ & 0.66 & 0.056\% & 0.53 & 0.57\% & 0.43 & 0.66\% \\
$M_{\rm P}$ & 0.45 & 3.1\% & 0.22 & 28\% & 0.23 & 15\% \\
$K$ & 0.50 & 1.5\% & 0.28 & 17\% & 0.28 & 8.8\% \\
$R_{\rm P}$ & -0.031 & 89\% & -0.053 & 80\% & -0.21 & 21\% \\
$M_{\rm S}$ & 0.19 & 39\% & 0.15 & 48\% & 0.30 & 6.8\% \\
$T_{\rm eq,P}$ & -0.19 & 38\% & -0.16 & 42\% & -0.19 & 24\% \\
$\langle F \rangle_{\rm P}$ & -0.21 & 33\% & -0.18 & 38\% & -0.21 & 19\% \\
$P$ & -0.29 & 17\% & -0.23 & 26\% & 0.0096 & 95\% \\
$[$Fe/H$]$ & -0.19 & 39\% & -0.18 & 38\% & -0.059 & 72\% \\
$T_{\rm eff,S}$ & -0.31 & 15\% & -0.32 & 11\% & -0.18 & 28\% \\
\enddata
\label{tab:correlationcoeffs}
\tablenotetext{a}{The Spearman nonparametric rank-order correlation coefficient}
\tablenotetext{b}{False alarm probability. These have not been corrected for the total number of independent trials conducted by searching for correlations between different parameter combinations.}
\end{deluxetable}

\begin{deluxetable}{lrrr}
\tabletypesize{\scriptsize}
\tablewidth{0pc}
\tablecaption{Adopted Values for the Stellar Activity Index and Planetary Surface Gravity}
\tablehead{
\multicolumn{1}{c}{Planet} &
\multicolumn{1}{c}{$\log{R^{\prime}_{\rm HK}}$\tablenotemark{a}} &
\multicolumn{1}{c}{$\log g_{P}$} &
\multicolumn{1}{c}{Ref. $\log g_{P}$} \\
& &
\colhead{[cgs]}
& 
}
\startdata
CoRoT-1b & -5.312 & 3.0266\tablenotemark{b} & 10 \\
CoRoT-2b & -4.331 & 3.6157\tablenotemark{b} & 1 \\
GJ 436b & -5.298 & 3.1070 & 8 \\
HAT-P-10/WASP-11b & -4.823 & 3.0800 & 16 \\
HAT-P-11b & -4.567 & 3.0500 & 23 \\
HAT-P-12b & -5.104 & 2.7500 & 19 \\
HAT-P-13b & -5.138 & 3.1088\tablenotemark{b} & 21,22 \\
HAT-P-14b & -4.855 & 3.6200 & 27 \\
HAT-P-1b & -4.984 & 2.9570 & 8 \\
HAT-P-2b & -4.780 & 4.2260 & 29 \\
HAT-P-3b & -4.904 & 3.3100 & 3 \\
HAT-P-4b & -5.082 & 3.0200 & 3 \\
HAT-P-5b & -5.061 & 3.2190 & 3 \\
HAT-P-6b & -4.799 & 3.1710 & 3 \\
HAT-P-7b & -5.018 & 3.3406\tablenotemark{b} & 4,5,30 \\
HAT-P-8b & -4.985 & 3.2300 & 18 \\
HAT-P-9b & -5.092 & 2.9910 & 12 \\
HD 149026b & -5.030 & 3.1320 & 17 \\
HD 17156b & -5.022 & 3.8810 & 15 \\
HD 189733b & -4.501 & 3.3099\tablenotemark{b} & 11 \\
HD 209458b & -4.970 & 2.9630 & 3 \\
HD 80606b & -5.061 & 4.0202\tablenotemark{b} & 28 \\
TrES-1b & -4.738 & 3.2200 & 3 \\
TrES-2b & -4.949 & 3.2980 & 3 \\
TrES-3b & -4.549 & 3.4250 & 13 \\
TrES-4b & -5.104 & 2.8580 & 13 \\
WASP-12b & -5.500 & 2.9900 & 14 \\
WASP-13b & -5.263 & 2.8500 & 9 \\
WASP-14b & -4.923 & 4.0100 & 24 \\
WASP-17b & -5.331 & 2.5600 & 26 \\
WASP-18b & -5.430 & 4.2810 & 20 \\
WASP-19b & -4.660 & 3.1900 & 25 \\
WASP-1b & -5.114 & 3.0100 & 3 \\
WASP-2b & -5.054 & 3.2870 & 3 \\
WASP-3b & -4.872 & 3.4200 & 2 \\
XO-1b & -4.958 & 3.2110 & 3 \\
XO-2b & -4.988 & 3.1680 & 3 \\
XO-3b & -4.595 & 4.2950 & 6 \\
XO-4b & -5.292 & 3.3316\tablenotemark{b} & 7 \\
\enddata
\label{tab:rhkvslogg}
\tablenotetext{a}{$\log R^{\prime}_{\rm HK}$ values are taken from KHI10.}
\tablenotetext{b}{Calculated from $K$, and a combination of $a/R_{\star}$, $T_{14}$, $\rho_{\star}$, $b$, $i$, and $R_{p}/R_{\star}$ from the given source(s).}
\tablerefs{1. \citet{2008A&A...482L..21A}; 2. \citet{2008A&A...492..603G}; 3. \citet{2008ApJ...677.1324T}; 4. \citet{2008ApJ...680.1450P}; 5. \citet{2010arXiv1001.0413W}; 6. \citet{2008ApJ...683.1076W}; 7. \citet{2008arXiv0805.2921M}; 8. \citet{2008MNRAS.386.1644S}; 9. \citet{2009A&A...502..391S}; 10. \citet{2009A&A...506..359G}; 11. \citet{2009A&A...506..377T}; 12. \citet{2009ApJ...690.1393S}; 13. \citet{2009ApJ...691.1145S}; 14. \citet{2009ApJ...693.1920H}; 15. \citet{2009ApJ...693..794W}; 16. \citet{2009ApJ...696.1950B}; 17. \citet{2009ApJ...696..241C}; 18. \citet{2009ApJ...704.1107L}; 19. \citet{2009ApJ...706..785H}; 20. \citet{2009ApJ...707..167S}; 21. \citet{2009ApJ...707..446B}; 22. \citet{2010arXiv1003.4512W}; 23. \citet{2009arXiv0901.0282B}; 24. \citet{2009MNRAS.392.1532J}; 25. \citet{2010ApJ...708..224H}; 26. \citet{2010ApJ...709..159A}; 27. \citet{2010arXiv1003.2211T}; 28. \citet{2010arXiv1004.0790H}; 29. \citet{2010MNRAS.401.2665P}; 30. \citet{2009ApJ...703L..99W}}
\end{deluxetable}

\section{Potential Selection Effects}\label{sec:selecteffect}

\begin{figure}[ht]
\epsscale{1.2} \plotone{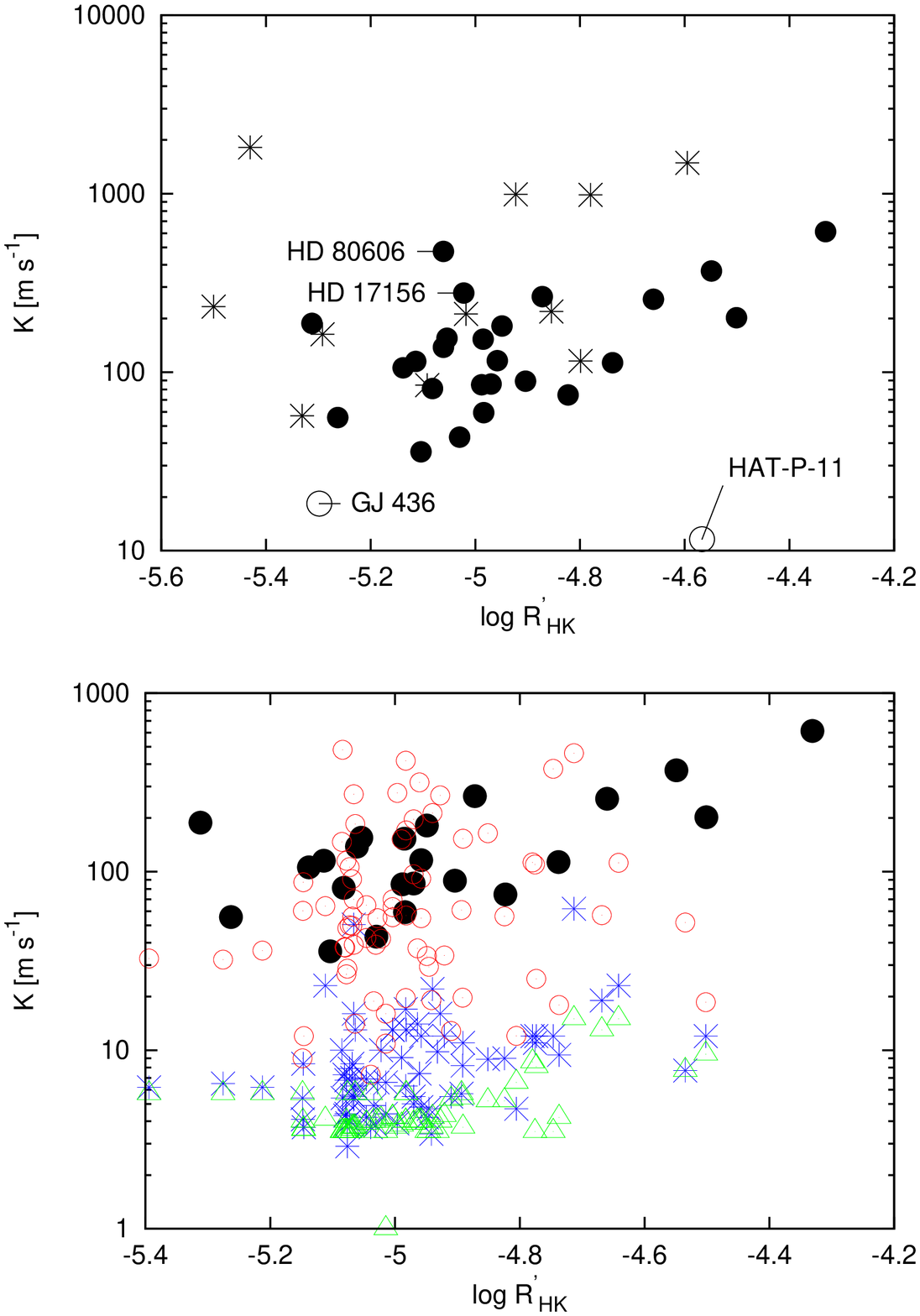}
\caption{Top: RV semiamplitude of TEPs vs. the chromospheric activity
  of the stellar hosts, as measured with $\log R^{\prime}_{\rm
    HK}$. Symbols are as in Figure~\ref{fig:rhkvslogg}. Bottom: Same
  as above, here we only show TEPs with $M > 0.1~M_{\rm J}$, $a <
  0.1~{\rm AU}$ orbiting stars with $4200~{\rm K} < T_{\rm eff} <
  6200~{\rm K}$ (filled black circles). We also show RV planets
  orbiting stars with $0.5 < (B-V) < 1.2$ from \citet{Butler.06} (red
  open circles) together with the residual RMS of an orbital fit to
  each RV planet (blue crosses), and the inferred stellar jitter
  (excess residual RMS not accounted for by the formal RV errors;
  green open triangles) for each RV planet. While the jitter and RMS
  for the RV planets does increase with stellar activity, the levels
  are still generally well below the $K$ values of the transiting hot
  Jupiters. Moreover, the RV planets themselves do not show a strong
  correlation between $K$ and $\log R^{\prime}_{HK}$, and indeed
  numerous RV planets have been discovered with $K < 30~{\rm
    m\,s^{-1}}$ and $\log R^{\prime}_{HK} > -5.0$.}
\label{fig:rhkvsK}
\end{figure}

The observed correlation between $\log R^{\prime}_{\rm HK}$ and $\log
g_{\rm P}$ may potentially be due to observational biases rather than
physical effects. The most obvious relevant observational bias is the
relative difficulty of obtaining high-precision RV observations for
high-activity stars. This might introduce a selection whereby lower
mass planets are not detectable if they are orbiting high activity
stars. Indeed, as seen in Table~\ref{tab:correlationcoeffs}, much of
the $\log R^{\prime}_{\rm HK}$-$\log g_{\rm P}$ correlation can be
attributed to a correlation between the RV semiamplitude $K$ and $\log
R^{\prime}_{\rm HK}$, with less contribution from the planet
radius. This is what would be expected for a selection effect of this
form. Figure~\ref{fig:rhkvsK}(top) shows the relation between these
two parameters. If this correlation were due only to an observational
bias, however, we would expect the correlation to continue to hold
when low mass and long period planets are also included (sample 2).
The fact that the correlation is reduced to $r_{S} = 0.28$ with a
false alarm probability of $17\%$ in this case is evidence that this
relation may not be due to a selection effect. The lowest $K$ planet
in this sample, HAT-P-11, was discovered around one of the most active
stars in the sample. Moreover, while a selection effect might explain
the absence of sample 1 planets in the lower right corner of
Figure~\ref{fig:rhkvsK}(top) or Figure~\ref{fig:rhkvslogg}(bottom), it
does not explain the absence of sample 1 planets in the upper left
corners of these figures, unless high $\log g_{\rm P}$ planets are
intrinsically less common than low $\log g_{\rm P}$ planets \emph{and}
low $\log R^{\prime}_{\rm HK}$ hot Jupiter host stars are intrinsically less
common than high $\log R^{\prime}_{\rm HK}$ host stars.

As an additional check on whether observational biases against
selecting planets with low $K$ around high activity stars could be
responsible for the $\log R^{\prime}_{\rm HK}$-$\log g_{\rm P}$
correlation, we compare the $\log R^{\prime}_{\rm HK}$-$K$ relation
for the TEPs from sample 1 to the sample of RV planets from the
California-Carnegie Planet Search presented by \citet{Butler.06}. We
consider here 84 planets from that sample with $S$ index measurements
and with $0.5 < (B-V) < 1.2$. We use the relations from
\citet{Noyes.84} to calculate $\log R^{\prime}_{\rm HK}$ from $(B-V)$
and $S$ for each of these stars. We also plot the jitter and residual
RMS from \citet{Butler.06}. While the residual RMS increases towards
higher stellar activity levels, it is generally well below $K =
30~{\rm m\,s^{-1}}$, and thus well below the lowest $K$ systems in
sample 1. Moreover, a significant number of RV planets with $K <
30~{\rm m\,s^{-1}}$ have been detected around stars with $\log
R^{\prime}_{\rm HK} > -5$. All this suggests that poorer RV precision
due to increased stellar activity is unlikely to explain the absence
of low surface gravity (or low $K$) hot Jupiters around active stars.

\section{Discussion}\label{sec:discussion}

Assuming that the observed correlation between $\log R^{\prime}_{\rm
  HK}$ and $\log g_{\rm P}$ is not due to an observational bias, it is
not obvious what physical processes might give rise to it. It is
well-known that $\log R^{\prime}_{\rm HK}$ is a decreasing function of
age for FGK stars \citep[e.g.][]{Soderblom.91}, but the $\log g_{\rm
  P}$-age relation implied from the $\log g_{\rm P}$-$\log
R^{\prime}_{\rm HK}$ relation is opposite of what is expected--that
planets should contract with age, and not expand with age. For
example, by interpolating the \citet{Fortney.07} models while
accounting for the increase in stellar luminosity over time, we find
that a coreless $1.0~M_{\rm J}$ planet orbiting a $1.0~M_{\odot}$ star
on a $2.5~{\rm day}$ period should decrease in radius from
$1.22~R_{\rm J}$ to $1.13~R_{\rm J}$ between 300~Myr and
4.5~Gyr. Models such as these, however, are known to underpredict the
radii of many hot Jupiters. If the effect of insolation on planetary
radii is substantially larger than anticipated, so that the inflation
due to the increase in stellar luminosity with time is greater than
the gravitational contraction of the planet over time, the result
would be a positive $\log g_{\rm P}$-$\log R^{\prime}_{\rm HK}$
correlation.

Another possibility is that strong stellar UV flux increases
the evaporation of hydrogen from the atmospheres of hot Jupiters,
leading to higher metallicity, more compact planets \citep[e.g.][and
references therein]{LecavelierdesEtangs.10}. If this were the case, we
might expect to see a correlation between planet radius and stellar
activity, and no correlation between planet mass and activity. The
fact that the opposite effect is observed casts doubt on this
hypothesis. Moreover, since the stellar activity should decrease with
age, this hypothesis does not explain how planets could re-inflate
when the activity is lowered.

Alternatively, the presence of hot Jupiters may induce activity on the
host star, either by tidally spinning-up the star's convection zone,
or via a magnetic star-planet interaction \citep[see the review
  by][]{Shkolnik.09}. In this case stars with high $\log
R^{\prime}_{\rm HK}$ may not necessarily be younger than stars with
lower $\log R^{\prime}_{\rm HK}$. Evidence that the presence of a hot
Jupiter is correlated with increased stellar X-ray activity has been
presented by \citet{Kashyap.08}, while \citet{Pont.09} found that hot
Jupiter host stars may exhibit excess rotation. Other investigations
have found evidence of magnetic activity variations correlated with
planet properties \citep[e.g.][]{Shkolnik.05,Lanza.09}. For both tidal
and magnetic star-planet interactions, the strength of the interaction
increases with planet mass. If the $\log R^{\prime}_{\rm HK}$-$\log
g_{\rm P}$ correlation is a by-product of a more fundamental $\log
R^{\prime}_{\rm HK}$-$M_{\rm P}$ correlation, one might wonder why the
former is detected with higher significance than the latter. A
possible explanation is that $\log g_{\rm P}$ is determined directly
from measurable parameters while $M_{\rm P}$ is directly proportional
to the stellar mass $M_{\rm S}$, which in turn is dependent on stellar
models. As a result $\log g_{\rm P}$ is generally determined with
better precision, and presumably with better accuracy, than $M_{\rm
  P}$ for TEPs. However, by simulating data sets with the observed
$M_{\rm P}$-$\log R^{\prime}_{\rm HK}$ correlation and $M_{\rm P}-R_{\rm
  P}$ correlations, assuming the scatter about these relations is
intrinsic, and assuming the observational errors for $M_{\rm P}$ and
$\log g_{\rm P}$ are realistic, we find that there is only a $\sim
1\%$ probability of the FAP of $\log g_{\rm P}$-$\log R^{\prime}_{\rm
  HK}$ being less than 0.1\% while the FAP of $M_{\rm P}$-$\log
R^{\prime}_{\rm HK}$ is greater than 1\%. Even if we assume the true
observational error on $M_{\rm P}$ is $\sim 0.5M_{\rm P}$, the
probability is only $\sim 6\%$. It is therefore unlikely that the
$\log R^{\prime}_{\rm HK}$-$M_{\rm P}$ relation is driving the $\log
R^{\prime}_{\rm HK}$-$\log g_{\rm P}$ relation.

In summary, we have identified a significant positive correlation between
stellar activity and planetary surface gravity. As far as we are aware
this correlation is unanticipated, and its cause is unclear.

\acknowledgements

I am greatly indebted to H.~Knutson, A.~Howard, and H.~Isaacson for
publishing the $\log R_{\rm HK}$ values on which this study is
based. I would also like to thank R.~Noyes, G.~Bakos and A.~Howard for
helpful discussions and comments on this paper. Support for this work
is through the HATNet project, via NASA grants NNG04GN74G, NNX08AF23G
and SAO IR\&D grants.

\end{document}